\documentclass[conference]{IEEEtran}

\usepackage{cite}
\usepackage{amsthm, amsmath,amssymb,amsfonts}
\usepackage{algorithmic}
\usepackage{graphicx}
\usepackage{pgfplots}
\usepackage[algoruled]{algorithm2e}
\usepackage{microtype}
\usepackage{tikz}
\usepackage{textcomp}
\usepackage{xcolor}
\usepackage{verbatim}
\usepackage[binary-units=true]{siunitx}
\DeclareSIUnit{\belmilliwatt}{Bm}
\DeclareSIUnit{\dBm}{\deci\belmilliwatt}
\def\BibTeX{{\rm B\kern-.05em{\sc i\kern-.025em b}\kern-.08em
		T\kern-.1667em\lower.7ex\hbox{E}\kern-.125emX}}

\begin{document}

	\newtheorem{proposition}{Proposition}	
	
	\title{Rate-Power Region of SWIPT Systems Employing Nonlinear Energy Harvester Circuits with Memory}
	
	\author{\IEEEauthorblockN{Nikita Shanin, Laura Cottatellucci, and Robert Schober}
		\IEEEauthorblockA{\textit{Friedrich-Alexander-Universit\"{a}t Erlangen-N\"{u}rnberg (FAU), Germany} }
	}
	
	\maketitle
	
	\begin{abstract}
		In this paper, we study the rate-power region of a simultaneous wireless information and power transfer (SWIPT) system where a transmitter (TX) broadcasts a common signal to an information receiver (IR) and an energy harvester (EH). Since practical EH circuits include a reactive element as part of their signal rectifier and the voltage on this element cannot drop or rise instantaneously, the EH circuit has \emph{memory}. We model the memory effect of the EH by a Markov reward chain. Furthermore, since an analytical model that includes all non-linear and memory effects of the EH circuit is not available, we employ a deep neural network (DNN) to model the Markov chain. We formulate an optimization problem to determine the rate-power region of the considered SWIPT system and propose an iterative algorithm based on sequential quadratic programming (SQP) to solve it. Our numerical results show that the optimal input distribution and the rate-power region depend on both the input power level at the EH and the symbol duration.
	\end{abstract}
	
	\section{Introduction}
	
	Due to the considerable growth of the number of low-power devices, the Internet-of-Things (IoT) has attracted significant attention in recent years. However, the problem of efficient recharging or replacing of the batteries of billions of IoT devices, such as wireless sensors, remains unsolved. A possible solution is to harvest energy from radio frequency (RF) signals. This prospect has fueled significant interest in simultaneous wireless information and power transfer (SWIPT) systems \cite{Varshney2008, Grover2010, Clerckx2018, Collado2014, Morsi2019, Varasteh2019}.
	
	SWIPT was studied first in \cite{Varshney2008}. The author showed that there exists a fundamental trade-off between the achievable information rate and the transferred power for discrete-time memoryless Gaussian channels. This trade-off can be characterized by a capacity-energy region. In \cite{Grover2010}, the authors showed that, in a frequency-selective channel with additive white Gaussian noise (AWGN), a simple sinusoidal signal is optimal for power transfer, whereas the waterfilling strategy is optimal for information transmission. 	 
		
    SWIPT systems employ an energy harvester (EH) to convert the received RF signal into a direct current (DC) signal. The EH includes a rectenna, i.e., an antenna followed by a rectifier. In \cite{Varshney2008} and \cite{Grover2010}, the authors assumed linear EH circuit models. However, recently, practical non-linear models for EH circuits were proposed for performance optimization of SWIPT systems \cite{Clerckx2018, Collado2014, Morsi2019, Varasteh2019, Varasteh2019b}. In \cite{Clerckx2018}, the author investigated a non-linear diode model obtained by a Taylor series approximation for the current flowing through the rectifier diode and showed that, for multi-carrier transmission, different input signal distributions maximize the information rate and the transferred energy, respectively. Thus, by varying the input distribution, different points of the rate-energy region can be achieved. Experiments in \cite{Collado2014} showed that signals with high peak-to-average power ratio (PAPR) yield larger harvested power compared to constant-envelope signals. In \cite{Morsi2019}, the authors developed a non-linear diode model and characterized the corresponding rate-energy region by optimally designing the input distribution which maximizes the mutual information between a transmitter (TX) and an information receiver (IR) for a given required harvested power value. In \cite{Varasteh2019}, based on the autoencoder concept and the EH circuit model in \cite{Morsi2019}, the authors proposed a learning approach to determine an error rate-power region by optimizing the modulation scheme. Finally, in \cite{Varasteh2019b}, employing a similar autoencoder concept, the authors adopted an EH circuit model similar to \cite{Clerckx2018} and a model based on a sigmoidal function for low and high EH input powers, respectively. The results in \cite{Varasteh2019b} suggest that On-Off signaling is optimal for power transfer, where for the low power regime, the probability of the On signal is small, and for the high power regime, the probability of the On signal is higher but its amplitude is smaller.
                
    The results in \cite{Clerckx2018, Morsi2019, Varasteh2019, Varasteh2019b} were obtained under strong assumptions regarding the EH circuits. In particular, it was assumed that the instantaneous harvested power depends on the currently received signal only. However, rectifier circuits typically include a reactive element (usually a capacitor) as part of a low-pass filter. Since the voltage (or current) level on this element cannot drop instantaneously \cite{Horowitz1989}, the rectenna circuit has \emph{memory}. Furthermore, for high RF signal powers, an EH exhibits the diode breakdown effect that was only partially included in \cite{Morsi2019} and completely neglected in \cite{Clerckx2018, Varasteh2019} and all other related works. Finally, the impedance values of the antenna and the rectifier have to be matched by a matching circuit which was assumed to be ideal in \cite{Clerckx2018, Morsi2019, Varasteh2019}. However, because of the rectifier non-linearity, perfect matching is possible for a single input signal frequency and a single power value only.
    	
	The goal of this paper is to analyze and design SWIPT systems taking into account the above mentioned effects, that were not properly addressed in the existing theoretical models. In practice, it is not possible to develop an analytical model for the EH circuit that includes all non-linear and memory effects. 
	Here, we model the memory of the EH by a Markov reward process. Thereby, we treat the output voltage levels as the states of a Markov reward chain and the amount of harvested power as the reward. 
	For this model, we propose an iterative algorithm based on sequential quadratic programming (SQP) \cite{Nocedal2006} for optimization of the input signal distribution.
	Then, we apply the framework proposed in \cite{Marbach2001} to obtain the direction of the gradient of the average reward with respect to the input signal distribution treating it as an underlying parameter of the Markov reward chain. 
	Additionally, we propose a learning approach to deal with the non-idealities of the EH circuit. 
	In particular, we utilize a dense neural network (DNN) to simulate the EH circuit and to predict the current reward given the state of the Markov reward chain and the received signal. 
	Our simulation results show that the optimal input distribution and the rate-power region depend on both the symbol duration and the input signal power of the EH. In particular, a shorter symbol duration increases the achievable bit rate at the expense of a decrease of the average harvested power. 
		
	The rest of the paper is organized as follows. In Section \ref{SystemModel_Section}, we introduce the system model, provide some background on the EH circuit, and define a Markov reward chain that models the energy harvesting process. In Section \ref{ProblemFormulation_Section}, we formulate the proposed optimization problem. In Section \ref{ProblemSolution_Section}, we develop an algorithm for solving the problem and design the DNN for EH circuit simulation. In Section \ref{SimulationResults_Section}, we provide simulation results for performance evaluation. Finally, in Section \ref{Conclusion_Section}, we draw some conclusions.
	
	Throughout this paper we use the following notations. Bold lower case letters stand for vectors, i.e., $\boldsymbol{x}$ is a vector, and its $i^\text{th}$ element is denoted by $\boldsymbol{x}_i$. The average value of a variable $x$ is denoted by $\overline{x}$. $f(x,y;z)$ denotes a function of variables $x$ and $y$ for a given parameter $z$. $f(x;z)\vert_{x={x_0}}$ is the value of function $f(x;z)$ at $x=x_0$. $\mathbb{E}_x\{\cdot\}$ denotes the expectation with respect to the distribution of random variable $x$. Operator $\Re\{\cdot\}$ denotes the real part of a complex number. $\left\lVert \cdot \right\rVert$ represents the Euclidean norm. $(\cdot)^\top$ denotes the transpose of a vector. $\mathbb{R}$ refers to the set of real numbers. The imaginary unit is denoted by $j$. The Gaussian distribution with mean $\mu$ and variance $\sigma^2$ is denoted by $\mathcal{N}(\mu, \sigma^2)$. $\mathrm{Pr} \{x = {x}_{i}\}$ denotes the probability that random variable $x$ is equal to a particular value ${x}_{i}$.

	\section{System Model and Preliminaries}
	
	\label{SystemModel_Section}
	
	\subsection{System Model}
\begin{figure}[t]
		\begin{center}
		\begin{tikzpicture}[scale=1]
		
		\draw (-.75,-.25) rectangle (.75,.25);
		\draw(0,0) node[align = center] {TX};		
		
		\draw (.75,0) -- (1.75,0);
		\draw (1.75,-0.5) -- (1.75,0.5);
		\draw(.8,0) node[align = left, above right] {$x(t)$};	
		
		\draw [->] (1.75,0.5) -- (1.95,0.5);
		\draw (1.95,0.2) rectangle (2.55,.8);
		\draw (2.25,0.5) node[align = center] {$h_I$};	
		\draw [->] (2.55,0.5) -- (3,0.5);
		\draw (3.25,0.5) circle (0.25);
		\draw (3.25,0.5) node[align = center] {+};	
		\draw [->] (3.25,1.25) -- (3.25,0.75);
		\draw (3.25,1.25) node[align = center, above] {$n(t)$};	
		
		\draw [->] (3.5,0.5) -- (4.5,0.5);
		\draw (3.5,0.5) node[align = left, above right] {$y_I(t)$};	
		
		\draw (4.5,0.25) rectangle (6,0.75);
		\draw(5.25,0.5) node[align = center] {IR};

		\draw [->] (1.75,-0.5) -- (2.8,-0.5);
		\draw (2.8,-0.2) rectangle (3.4,-.8);
		\draw (2.8+0.3,-0.5) node[align = center] {$h_E$};
		\draw [->] (3.4,-.5) -- (4.5,-0.5);
		\draw (3.5,-0.5) node[align = left, above right] {$y_E(t)$};
		
		\draw (4.5,-0.25) rectangle (6,-0.75);
		\draw(5.25,-0.5) node[align = center] {EH};		
		
		\end{tikzpicture}
		\end{center}
		
		\caption{SWIPT system model comprising a transmitter (TX), an information receiver (IR), and an energy harvester (EH).}
		\label{System_Fig}
		\end{figure}
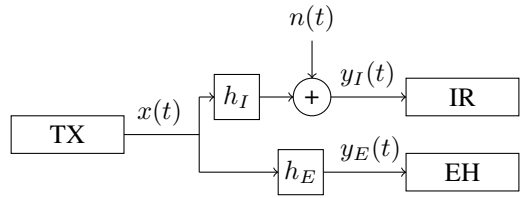

	Let us consider the SWIPT system in Fig. \ref{System_Fig}. It consists of three nodes: a TX, an IR, and an EH. The TX broadcasts a pulse-modulated signal, which is received by both the IR and EH. 
	This signal is modeled as $x(t) = \sum_{k=0}^{\infty}x[k] \psi (t-kT)$, where $T$ is the symbol duration, $\psi(t)$ is the transmit pulse shape, and $x[k]$ are the information symbols taken from a real-valued\footnote{As is customary for information theoretical analysis, see e.g. \cite{Cover2012}, for the sake of clarity, we assume a real-valued constellation set. A further generalization to a complex constellation is relatively straightforward but omitted here due to space constraints.} constellation set $\mathcal{X} \subset \mathbb{R}$ of size $S$.   
	
	The symbols $x[k], k \in \{0, 1, ...\},$ are random variables with discrete probability mass function (pmf) $p_x(x)$, modeled by a vector $\boldsymbol{\theta} \in [0,1]^S$. 
	Here, $\boldsymbol{\theta}_i$ is the probability that random variable $x[k]$ takes the $i^\text{th}$ value, $i \in \{0, 1, ..., S-1\}$, from constellation set $\mathcal{X}$. 
	The channel gains of the IR and EH are assumed to be perfectly known and are denoted by $h_I \in \mathbb{R}$ and $h_E \in \mathbb{R}$, respectively. 
	Hence, the RF signals received at the IR and EH can be expressed as $y_{I}^{RF}(t) = \sqrt{2}\Re\{[h_I x(t) + n(t)] e^{j 2 \pi f_ct}\}$ and $y_{E}^{RF}(t) = \sqrt{2}\Re\{h_E x(t) \, e^{j 2 \pi f_ct}\}$, where $f_c$ and $n(t)$ denote the carrier frequency and real-valued zero-mean AWGN, respectively. 
	We note that the noise received at the EH is ignored because its contribution to the harvested energy is negligible.

	\subsection{Information Receiver}
	
	Let us consider signal $y_{I}^{RF}(t)$ received at the IR. 
	Since $y_{I}^{RF}(t)$ is a time-slotted signal, after down-conversion, matched filtering, and sampling, the received signal in time interval $k$ can be expressed as $y[k] = h_I x[k] + n[k]$, where $y[k]$ is the information channel output following probability density function (pdf) $p_y(y)$ and $n[k]$ is the discrete-time zero-mean AWGN with variance $\sigma^2_n$.
	
	The mutual information between  $x[k]$ and $y[k]$ as a function of the input distribution $\boldsymbol{\theta}$ can be expressed as \cite{Smith1971} $I(\boldsymbol{\theta}) = H_y(\boldsymbol{\theta}) - H_n$, where $H_y(\boldsymbol{\theta})$ and $H_n$ are the differential entropies of the received signal and the noise, respectively. 
	The differential entropy of the noise does not depend on input distribution $\boldsymbol{\theta}$ and is equal to $H_n = \frac{1}{2}\text{log}_2(2 \pi e \sigma_n^2)$. 
	The differential entropy of the received signal is given by \cite{Cover2012} $H_y(\boldsymbol{\theta}) = -\int_{y} p_y(y; \boldsymbol{\theta}) \, \text{log}_2 \, \big( p_y(y; \boldsymbol{\theta}) \big) dy $. 
	Since $y[k]$ is a sum of independent and identically distributed (i.i.d.) random variables, $p_y(y)$ can be obtained as convolution of the individual distributions, i.e., $p_y(y)  = \sum_{i} \mathrm{Pr}  \{{x} = {x}_{i}\} \linebreak \times p_n(n=y-h_I{x}_{i}) = \sum_{i} \boldsymbol{\theta}_i \, p_n(n=y-h_I{x}_i).$ 
	Therefore, the output differential entropy measured in bits per symbol is given by	
	
	\begin{align}
	H_y(\boldsymbol{\theta}) = - \int_{y} \sum_{i} \boldsymbol{\theta}_i \, p_n&(y-h_I{x}_i) \nonumber\\
	&\log_2 \big( \sum_{i} \boldsymbol{\theta}_i \, p_n(y-h_I{x}_i) \big) dy.
	\label{MutInf_Eqn}
	\end{align}

	\subsection{EH Circuit}
	\label{EHCircuit_Section}

	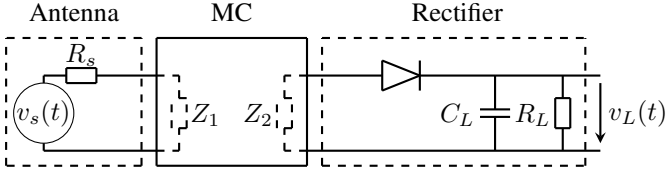
\begin{figure}[t]
		\begin{center}		
			\begin{tikzpicture}		
			
			\draw[thick,rounded corners=8pt] (-0.2, 0.9) -- (-0.2, 1.1);
			\draw[thick,rounded corners=8pt] (0.2, 0.9) -- (0.2, 1.1);
			\draw[thick,rounded corners=8pt] (0.2, 1.1) -- (-0.2, 1.1);
			\draw[thick,rounded corners=8pt] (0.2, 0.9) -- (-0.2, 0.9);
			\draw(0, 1) node[align = center, above] {$R_s$};
			
			\draw(1.65, 0.2) node[align = center, above] {$Z_1$};
			\draw[thick,rounded corners=8pt, dashed] (1, 0) -- (1.3, 0);
			\draw[thick,rounded corners=8pt, dashed] (1, 1) -- (1.3, 1);
			\draw[thick,rounded corners=8pt, dashed] (1.3, 0) -- (1.3, 0.3);
			\draw[thick,rounded corners=8pt, dashed] (1.4, 0.3) -- (1.4, 0.7);
			\draw[thick,rounded corners=8pt, dashed] (1.2, 0.3) -- (1.2, 0.7);
			\draw[thick,rounded corners=8pt, dashed] (1.4, 0.3) -- (1.2, 0.3);
			\draw[thick,rounded corners=8pt, dashed] (1.4, 0.7) -- (1.2, 0.7);
			\draw[thick,rounded corners=8pt, dashed] (1.3, 0.7) -- (1.3, 1);

			\draw(2.35, 0.2) node[align = center, above] {$Z_2$};
			\draw[thick,rounded corners=8pt, dashed] (3, 0) -- (2.7, 0);
			\draw[thick,rounded corners=8pt, dashed] (3, 1) -- (2.7, 1);
			\draw[thick,rounded corners=8pt, dashed] (2.7, 0) -- (2.7, 0.3);
			\draw[thick,rounded corners=8pt, dashed] (2.8, 0.3) -- (2.8, 0.7);
			\draw[thick,rounded corners=8pt, dashed] (2.6, 0.3) -- (2.6, 0.7);
			\draw[thick,rounded corners=8pt, dashed] (2.8, 0.3) -- (2.6, 0.3);
			\draw[thick,rounded corners=8pt, dashed] (2.8, 0.7) -- (2.6, 0.7);
			\draw[thick,rounded corners=8pt, dashed] (2.7, 0.7) -- (2.7, 1);
			
			\draw(2, 1.6) node[align = center, above] {MC};

			\draw (-0.5, 0.5) circle (0.4);
			\draw(-0.5, 0.5) node[align = center] {$v_s(t)$};
			
			\draw[thick,rounded corners=8pt] (-0.2, 1) -- (-0.5, 1);
			\draw[thick,rounded corners=8pt] (-0.5, 0) -- (-0.5, 0.1);
			\draw[thick,rounded corners=8pt] (-0.5, 0.9) -- (-0.5, 1);
			\draw[thick,rounded corners=8pt] (-0.5, 0) -- (1, 0);
			\draw[thick,rounded corners=8pt] (0.2, 1) -- (1, 1);
			
			\draw[thick,rounded corners=8pt] (1, -0.2) -- (3, -0.2);
			\draw[thick,rounded corners=8pt] (1, 1.5) -- (3, 1.5);
			\draw[thick,rounded corners=8pt] (1, -0.2) -- (1, 1.5);
			\draw[thick,rounded corners=8pt] (3, -0.2) -- (3, 1.5);

			\draw[thick,rounded corners=8pt] (3, 1) -- (4, 1);
			\draw[thick,rounded corners=8pt] (3, 0) -- (6.9, 0);
			\draw[thick,rounded corners=8pt] (4, 1.2) -- (4, 0.8);
			\draw[thick,rounded corners=8pt] (4, 1.2) -- (4.5, 1);
			\draw[thick,rounded corners=8pt] (4, 0.8) -- (4.5, 1);
			\draw[thick,rounded corners=8pt] (4.5, 1.2) -- (4.5, 0.8);
			\draw[thick,rounded corners=8pt] (4.5, 1) -- (5.5, 1);
			
			\draw[thick,rounded corners=8pt] (5.5, 0.45) -- (5.5, 0);
			\draw[thick,rounded corners=8pt] (5.7, 0.45) -- (5.3, 0.45);
			\draw[thick,rounded corners=8pt] (5.7, 0.55) -- (5.3, 0.55);
			\draw[thick,rounded corners=8pt] (5.5, 1) -- (5.5, 0.55);
			
			\draw(5, 0.2) node[align = center, above] {$C_L$};
			
			\def\x{6.4}
			\draw[thick,rounded corners=8pt] (\x, 0.3) -- (\x, 0);
			\draw[thick,rounded corners=8pt] (\x, 1) -- (\x, 0.7);
			
			\draw[thick,rounded corners=8pt] (\x+0.1, 0.3) -- (\x+0.1, 0.7);
			\draw[thick,rounded corners=8pt] (\x-0.1, 0.3) -- (\x-0.1, 0.7);
			\draw[thick,rounded corners=8pt] (\x+0.1, 0.3) -- (\x-0.1, 0.3);
			\draw[thick,rounded corners=8pt] (\x+0.1, 0.7) -- (\x-0.1, 0.7);
			
			\draw(6, 0.2) node[align = center, above] {$R_L$};
			
			\draw[thick,rounded corners=8pt] (5.5,1) -- (\x+.5,1);		
			
			\draw[thick,rounded corners=8pt, stealth-] (6.9, 0.1) -- (6.9, 0.9);
			\draw(7.4, 0.2) node[align = center, above] {$v_L(t)$};
			
			\draw(5, 1.6) node[align = center, above] {Rectifier};
			\draw[dashed, thick, rounded corners=8pt] (6.7, -0.2) -- (6.7, 1.5);
			\draw[dashed, thick, rounded corners=8pt] (3.2, -0.2) -- (3.2, 1.5);
			\draw[dashed, thick, rounded corners=8pt] (3.2, -0.2) -- (6.7, -0.2);
			\draw[dashed, thick, rounded corners=8pt] (3.2, 1.5) -- (6.7, 1.5);

			\draw(-0.1, 1.6) node[align = center, above] {Antenna};
			\draw[dashed, thick, rounded corners=8pt] (-1,-0.2) -- (-1, 1.5);
			\draw[dashed, thick, rounded corners=8pt] (0.8, -0.2) -- (0.8, 1.5);
			\draw[dashed, thick, rounded corners=8pt] (0.8, -0.2) -- (-1, -0.2);
			\draw[dashed, thick, rounded corners=8pt] (0.8, 1.5) -- (-1, 1.5);
			\end{tikzpicture} 		
		\end{center}
		
		\caption{EH circuit model comprising an antenna, a matching circuit (MC), and a rectifier.}
		\label{EH_Fig}	
	\end{figure}
	
	Similar to \cite{Morsi2019} and references therein, we assume that the EH is equipped with a rectenna circuit as shown in Fig. \ref{EH_Fig}. The antenna is modeled as a voltage source $v_s(t)$ connected in series with resistance $R_s$. The rectifier circuit consists of a diode, a low-pass RC filter composed of the diode resistance and a capacitor $C_L$, and a load resistor $R_L$. Thus, the RF signal received at the EH node $y_{E}^{RF}(t)$ is converted by the rectenna circuit to a low frequency output voltage $v_L(t)$ across the load resistance $R_L$. Additionally, as in \cite{Morsi2019} and \cite{LePolozec2016}, we include an impedance matching circuit (MC) to maximize the power transferred from the antenna to the rectifier. Thus, we match antenna output impedance $Z_1$ and the input impedance of the rectifier circuit $Z_2$. Note that since the circuit includes a non-linear element, namely the diode, exact matching is possible for one frequency and one power value of the received signal only.
	
	The instantaneous harvested power can be expressed as ${P(t)} =  \frac{v_L^2(t)}{R_L}$. 
	Since the received RF signal $y_{E}^{RF}(t)$ is time-slotted, so is the output voltage signal $v_L(t)$, and hence, the harvested power $P(t)$. Moreover, due to the presence of the low-pass filter, in every time interval $k$, the signal $v_{L}(t)$, $t \in [(k-1)T, kT]$, depends not only on the transmitted symbol $x[k]$, but also on the output voltage level at the end of the previous time slot, $v_{L}((k-1)T)$. Hence, the rectenna has \emph{memory}.
	
	In general, $P(t)$ is a random process. We denote the average power harvested during the transmission of an infinitely long sequence of random symbols $\{x[k]\}$ by $\overline{P}$. This value can be estimated by averaging function $P(t)$ over time, or equivalently, over time intervals, assuming that the number of time intervals $K$ approaches infinity:
	
	\setlength{\arraycolsep}{0.0em}
	\begin{eqnarray}
	&\overline{P} &= \lim_{t \rightarrow \infty} \frac{1}{t} \int_{0}^{t} P(\tau) d\tau = \nonumber\\
	&&= \lim_{K \rightarrow \infty} \frac{1}{K} \sum_{k=0}^{K-1} \frac{1}{T} \int_{0}^{T} P(t+kT) dt.
	\label{EHCircuitEqn}
	\end{eqnarray}
	\setlength{\arraycolsep}{5pt}

	\subsection{Markov Chain Model of EH Circuit}
	\label{MarkovChainSection}
	
	Since $v_L(t)$ is a stochastic process, in the following, we employ the concept of Markov reward chain \cite{Howard1971} to model it. 
	In particular, we treat the load voltages at the end of the symbol intervals as the states of a Markov chain. 
	Moreover, we define a reward associated with each state such that the average power in (\ref{EHCircuitEqn}) coincides with the average reward of the Markov chain.
	
	Let us map the voltage levels $v_L$ to the states $\xi \in \Xi$ of a stochastic process, where $\Xi$ is a continuous state space.
	Thus, the system is in state $\xi[k]=\xi'$, if at time instant $t=kT$ the load voltage level equals the value associated with this state, i.e., $\xi[k] = \xi' = v_L'= v_L\big(kT\big)$. 
	Note that the voltage level at the load resistance is always bounded because of the diode breakdown effect \cite{Guo2014}.
	The discrete-time stochastic process $\{\xi[k]\}$ may change its value in each symbol interval, i.e., when the EH receives a new symbol. Hence, $\{\xi[k]\}$ is a discrete-time process and its \emph{time step} is equal to the symbol duration.	
	The behavior of the EH circuit in a given time interval is completely determined by the initial conditions and the input signal \cite{Horowitz1989}. 
	Hence, $\xi[k]=v_L(kT)$ depends only on the voltage level of the load resistance at time $(k-1)T$ and the received signal. 
	Thus, the probability of any state of the chain $\xi[k]$ depends only on the previous state $\xi[k-1]$, i.e., $\mathrm{Pr} \{ \xi[k]\,|\,\xi[{k-1}],\,\xi[{k-2}],\,...,\,\xi[{0}] \}~=~\mathrm{Pr} \{ \xi[k]\,|\,\xi[k-1]\}$ and $\{\xi[k]\}$ can be modeled by a Markov chain \cite{Norris1998, Meyn2012}.
	
	The transition probabilities of this Markov chain depend on $\boldsymbol{\theta}$. We denote the transition pdf from state $\xi_\mu \in \Xi$ to state $\xi_\nu \in \Xi$ by $\rho(\xi_\mu,\xi_\nu;\boldsymbol{\theta})$. The pdf $\rho(\xi_\mu, \xi_\nu;\boldsymbol{\theta})$ is non-zero if and only if there exists a symbol ${x}_{i} \in \mathcal{X}$, which occurs with non-zero probability $\boldsymbol{\theta}_i$, such that reception of $h_E{x}_{i}$ leads to a transition from $\xi_\mu=\xi'$ to $\xi_\nu=\xi''$. Moreover, in this case, the transition probability from $\xi'$ to $\xi''$ is equal to $\rho(\xi_\mu, \xi_\nu;\boldsymbol{\theta}) \vert_{\xi_\mu=\xi', \xi_\nu=\xi''} = \boldsymbol{\theta}_i$. Thus, the pdf $\rho(\xi_\mu,\xi_\nu;\boldsymbol{\theta})$ is differentiable with respect to $\boldsymbol{\theta}$ for any pair of states $\{\xi_\mu,\xi_\nu\}$. In the following, we are interested only in pairs of states for which a transition $\xi_\mu \rightarrow \xi_\nu$ exists.
	
	Since $\{\xi[k]\}$ is a Markov chain, given any state $\xi[k]$, a random sequence of received symbols $\{h_Ex[k], h_Ex[k+1], ... \}$ generates a sequence of states $\{\xi[k], \xi[k+1], \xi[k+2],... \}$, refered to as \emph{random walk}, which is a possible realization of the Markov chain starting from state $\xi[k]$. 
	We note that the analytical computation of state $\xi[k+1]$ given the current state $\xi[k]$ does not seem tractable due to the diode non-linearity, the imperfections of the matching network, and the circuit memory. Therefore, in Section \ref{NN_Section}, we will employ a DNN \cite{Goodfellow2016} to estimate $\xi[k+1]$  for given $\xi[k]$ and $x[k]$.
	
	During every transition of the Markov chain, the amount of power harvested by the EH depends on the received symbol $h_E{x}[k]$ and on the previous state $\xi[k]=\xi_n$. 
	We include this amount of power to the reward attained when the Markov chain is in state $\xi_n$ \cite{Howard1971}. 
	To define the reward at state $\xi$, let us consider the $\emph{average harvested power}$ $\overline{P}$ in (\ref{EHCircuitEqn}). Since the result of the integration  $\frac{1}{T}$$\int_{0}^{T} P(t+kT) dt$ depends on the current Markov chain state $\xi[k]={v_L}(kT)$ and the received symbol $h_E{x}[k]$ only, which are mutually statistically independent realizations of random variables $\xi$ and ${x}$, respectively, we define the average harvested power at state $\xi[k]$ corresponding to the received symbol $h_Ex[k]$ as $P'(\xi[k], h_E{x}[k]) = \frac{1}{T} \int_{0}^{T} P(t+kT) dt$. 
	Additionally, we assume that the Markov reward chain is ergodic, i.e., starting from any initial state, the Markov chain reaches the same limiting distribution \cite{Norris1998}. 
	Under this assumption, we can neglect the influence of the initial state and determine the average harvested power as follows 
	
	\begin{align}
	\overline{P} = \lim_{K \rightarrow \infty} \frac{1}{K}\Big[ \sum_{k=0}^{K-1} P'(\xi[k], h_E{x}[k])\Big] = \nonumber \\
	= \mathbb{E}_{\xi} \big\{ \mathbb{E}_{{x}} \big\{ P'(\xi, h_E{x}) \big\} \big\}.
	\label{AHPTimeToRVS__Eqn}
	\end{align}
	
	Expression (\ref{AHPTimeToRVS__Eqn}) suggests to define $\mathbb{E}_{{x}} \big\{ P'(\xi, h_E{x}) \big\}$ as the average harvested power attained under steady state conditions during a transition starting in state $\xi$. Then, value $\overline{P}$ can be interpreted as the average reward, obtained by performing a \emph{random walk}. 
	For a fixed constellation set $\mathcal{X}$ of size $S$, let $\boldsymbol{p}(\xi) \in \mathbb{R}^S$ be the vector of the possible values of $P'(\xi, h_E{x_i})$, i.e., its $i^\text{th}$ element, $\boldsymbol{p}_i(\xi)=P'(\xi, h_E{x_i})$, is the power attained when the Markov chain is in state $\xi$ and signal $h_E\,x_i$ is received. Since $\boldsymbol{p}(\xi)$ is also not analytically tractable, in Section \ref{NN_Section}, we employ a second DNN to estimate the elements of this vector.
	Since the pmf of the transmitted symbols is characterized by vector $\boldsymbol{\theta}$, the reward associated with state $\xi$ can be calculated as $\mathbb{E}_{{x}} \big\{ P'(\xi, h_E{x}) \big\} = \boldsymbol{\theta}^\top \boldsymbol{p}(\xi)$.
	
	Finally, we denote the limiting distribution of states $\xi$ in the steady state by $\pi(\xi,\boldsymbol{\theta})$. This distribution depends on $\boldsymbol{\theta}$ as a unique solution of the balance system of equations \cite{Meyn2012}
	
	\begin{equation}
	\int_{\xi_\mu} \pi(\xi_\mu;\boldsymbol{\theta}) \rho(\xi_\mu,\xi_\nu;\boldsymbol{\theta})d\xi_\mu = \pi(\xi_\nu; \boldsymbol{\theta})
	\label{BalanceEqn1}
	\end{equation}
	
	\begin{equation}	
	\int_{\xi} \pi(\xi; \boldsymbol{\theta})d\xi = 1
	\label{BalanceEqn2}
	\end{equation}	
		 
	Thus, from (\ref{AHPTimeToRVS__Eqn}) we obtain
	
	\setlength{\arraycolsep}{0.0em}
	\begin{eqnarray}
	&\overline{P} \equiv \overline{P}(\boldsymbol{\theta}) \, &= \boldsymbol{\theta}^\top \, \int_{\xi}\pi(\xi;\boldsymbol{\theta}) \boldsymbol{p}(\xi) d\xi.
	\label{ExpandedAverageRewardEqn}
	\end{eqnarray}
	\setlength{\arraycolsep}{5pt}

	\section{Optimization Problem Formulation}
	\label{ProblemFormulation_Section}
	
	In this section, we formulate an optimization problem to obtain the rate-power region of the considered SWIPT system. 

	In the following, we refer to the set of all attainable pairs of average harvested powers and achievable rates as the rate-power region of the SWIPT system. The boundary of this rate-power region for a given symbol duration $T$ can be obtained by solving the following constrained optimization problem:
	
	\begin{subequations}
		\label{MajorOptProblem}
		\begin{align}
		\label{MajorOptProblem_Objective}
		\max_{\boldsymbol{\theta}} \; &\overline{P}(\boldsymbol{\theta})\\
		\label{MajorOptProblem_Constr1}
		\text{subject to}\quad   &I(\boldsymbol{\theta}) \geq I_{\text{req}}, \\
		\label{MajorOptProblem_Constr2}
		&\mathbb{E}_x\{x^2\} \leq \sigma^2_x, \\
		\label{MajorOptProblem_Constr3}
		&\sum_{i=1}^{S} \boldsymbol{\theta}_i = 1,
		\end{align}
	\end{subequations}
	
	\noindent	
	where we maximize the power harvested at the EH (\ref{MajorOptProblem_Objective}) subject to a minimum required mutual information $I_{\text{req}}$ between TX and IR (\ref{MajorOptProblem_Constr1}). Constraint (\ref{MajorOptProblem_Constr2}) limits the average power (AP) budget at the TX to $\sigma^2_x$. Peak power constraints at the TX are incorporated by the proper choice of the constellation set $\mathcal{X}$. Constraint (\ref{MajorOptProblem_Constr3}) ensures that $\boldsymbol{\theta}$ corresponds to a valid pmf.			
	
	The problem in (\ref{MajorOptProblem}) is a non-convex optimization problem since the objective function in (\ref{MajorOptProblem_Objective}) is not concave with respect to $\boldsymbol{\theta}$. Thus, determining the globally optimal solution entails a high computational complexity. Therefore, in the following, we propose a low-complexity iterative algorithm to obtain a suboptimal solution of (\ref{MajorOptProblem}).
	
	\section{Solution of the Optimization Problem}
	\label{ProblemSolution_Section}
	\label{OptAlgSection}
	
	In this section, we design an algorithm based on sequential quadratic programming (SQP) \cite{Nocedal2006} to solve optimization problem (\ref{MajorOptProblem}). 
	To this end, we first formulate a convex quadratic subproblem that will be solved in each iteration of the proposed algorithm. 
	Then, we determine the gradients of all the functions included in (\ref{MajorOptProblem}) as they are needed to solve the formulated subproblem. Furthermore, we propose an iterative algorithm to obtain a suboptimal solution of (\ref{MajorOptProblem}). 
	Finally, we discuss the learning approach used in the proposed algorithm to simulate the Markov reward chain modeling the EH circuit.
			
	\subsection{Quadratic Subproblem}
	Since the objective function and the constraints in (\ref{MajorOptProblem}) are differentiable functions, in the following, we employ the SQP method \cite{Nocedal2006} to design an iterative algorithm that yields a suboptimal solution of (\ref{MajorOptProblem}). This method is known for its low complexity, fast convergence speed, and high efficiency in obtaining a local optimal solution of non-linear constrained optimization problems. Adapting the SQP approach from \cite{Nocedal2006} to problem (\ref{MajorOptProblem}), at iteration $k$ and for the current distribution $\boldsymbol{\theta}^k$, we obtain a search direction $\boldsymbol{\delta}^*$ as solution of a quadratic subproblem which approximates (\ref{MajorOptProblem}) as
	\begin{subequations}
		\label{OptSubproblem}
		\begin{align}
		\max_{\boldsymbol{\delta}} \overline{P}(\boldsymbol{\theta}^k) &+ \nabla_{\boldsymbol{\theta}}^\top \overline{P}(\boldsymbol{\theta}^k) \boldsymbol{\delta} + \frac{1}{2}\boldsymbol{\delta}^\top   \mathcal{H}^k \boldsymbol{\delta}   \\
		\text{subject to}\quad   &\nabla_{\boldsymbol{\theta}}^\top I(\boldsymbol{\theta}^k) \boldsymbol{\delta} +I(\boldsymbol{\theta}^k)  \geq I_{\text{req}},  \\
		&\nabla_{\boldsymbol{\theta}}^\top \mathbb{E}_x\{x^2\} \boldsymbol{\delta} + \mathbb{E}_x\{x^2\} \leq \sigma^2_x, \\
		&\sum_{i=1}^{S} \big( \boldsymbol{\theta}^k_i + \boldsymbol{\delta}_i \big) = 1,		
		\end{align}
	\end{subequations}
	\noindent
	where $\boldsymbol{\delta} \in \mathbb{R}^S$, $\mathcal{H}^k = \mathcal{H}(\boldsymbol{\theta}^k, \boldsymbol{\lambda}^k)$ is the Hessian matrix of the Lagrangian $\mathcal{L}(\boldsymbol{\theta}, \boldsymbol{\lambda})$ of the original problem (\ref{MajorOptProblem}) in the $k^\text{th}$ iteration, and $\boldsymbol{\lambda}^k$ is the corresponding vector of Lagrangian multipliers. 
	The Lagrangian of (\ref{MajorOptProblem}) is given by $\mathcal{L}(\boldsymbol{\theta}, \boldsymbol{\lambda}) = \overline{P}(\boldsymbol{\theta}) - \boldsymbol{\lambda}_1 \big(I(\boldsymbol{\theta}) - I_\text{req} \big) + \boldsymbol{\lambda}_2 \big(\mathbb{E}_x\{x^2\} - \sigma_x^2\big) + \boldsymbol{\lambda}_3 \big(\sum_{i=1}^{S} \boldsymbol{\theta}_i - 1 \big)$, where
	$\boldsymbol{\lambda}_i$ is the $i^\text{th}$ element of the vector of Lagrangian multipliers $\boldsymbol{\lambda} = (\boldsymbol{\lambda}_1, \boldsymbol{\lambda}_2, \boldsymbol{\lambda}_3)^\top$.
		 
	The system parameter $\boldsymbol{\theta}$ and the Lagrangian multipliers $\boldsymbol{\lambda}$ for the next iteration of the algorithm are updated as $\boldsymbol{\theta}^{k+1} = \boldsymbol{\theta}^k + \boldsymbol{\delta}^*$ and $\boldsymbol{\lambda}^{k+1} = \boldsymbol{\zeta}^*$.
	Here, $\boldsymbol{\delta}^*$ and $\boldsymbol{\zeta}^*=(\boldsymbol{\zeta}_1^*, \boldsymbol{\zeta}_2^*, \boldsymbol{\zeta}_3^*)^\top$ are the solution of (\ref{OptSubproblem}) and the vector of corresponding Lagrangian multipliers, respectively. 
	Note that (\ref{OptSubproblem}) is a convex quadratic optimization problem that can be solved by a numerical solver such as CVX \cite{Grant2015} provided that $\mathcal{H}^k$ and the gradients $\nabla_{\boldsymbol{\theta}} \overline{P}(\boldsymbol{\theta}^k)$, $\nabla_{\boldsymbol{\theta}} I(\boldsymbol{\theta}^k)$, and $\nabla_{\boldsymbol{\theta}} \mathbb{E}_x\{x^2\}$ are known.
	In each iteration, we update the Hessian matrix $\mathcal{H}^k$ using the Broyden-Fletcher-Goldfarb-Shanno (BFGS) method \cite{Nocedal2006} as follows:
	
	\begin{align}
		\label{BFGS}
			&\mathcal{H}^{k+1} = \mathcal{H}^k - \frac{\mathcal{H}^k \boldsymbol{\delta}^{*} \boldsymbol{\delta}^{* \top} \mathcal{H}^{k \top}}{\boldsymbol{\delta}^{* \top} \mathcal{H}^k \boldsymbol{\delta}^{*}} + \frac{\boldsymbol{y}^k \boldsymbol{y}^{k \top}}{\boldsymbol{y}^{k \top} \boldsymbol{\delta}^{*}},					
	\end{align}
	\noindent
	where $\boldsymbol{y}^k = \nabla_{\boldsymbol{\theta}} \mathcal{L}^{k+1}	- \nabla_{\boldsymbol{\theta}} \mathcal{L}^{k}$. 
	Here, $\mathcal{L}^k = \mathcal{L} (\boldsymbol{\theta}^k, \boldsymbol{\lambda}^k)$ denotes the Lagrangian in the $k^\text{th}$ iteration, i.e.,  $\nabla_{\boldsymbol{\theta}} \mathcal{L}^k = \nabla_{\boldsymbol{\theta}} \overline{P}(\boldsymbol{\theta}^k) + \boldsymbol{\lambda}^k_1 \nabla_{\boldsymbol{\theta}} I(\boldsymbol{\theta}^k) + \boldsymbol{\lambda}^k_2 \nabla_{\boldsymbol{\theta}}  \mathbb{E}_x\{x^2\} + \boldsymbol{\lambda}^k_3 \boldsymbol{1}_S$, where $\boldsymbol{1}_S$ is the all-ones column vector of size $S$.	
	
	Note that SQP with BFGS approximation converges superlinearly to a local optimum $\boldsymbol{\theta}^*$ if the initial points $\boldsymbol{\theta}^1$ and $\mathcal{H}^1$ are chosen such that $\left\lVert \boldsymbol{\theta}^1 - \boldsymbol{\theta}^*\right\rVert$ and $\left\lVert \mathcal{H}^1 - \nabla_{\boldsymbol{\theta} \boldsymbol{\theta}}^2 \mathcal{L}(\boldsymbol{\theta}^*, \boldsymbol{\lambda}^*) \right\rVert$ are sufficiently small, where $\nabla_{\boldsymbol{\theta} \boldsymbol{\theta}}^2 \mathcal{L}(\boldsymbol{\theta}^*, \boldsymbol{\lambda}^*)$ is the Hessian matrix of the Lagrangian for the optimal point \cite{Norris1998}.
	
	\subsection{Iterative Algorithm Design}	
	In the following, we first provide the gradients of constraints (\ref{MajorOptProblem_Constr1}), (\ref{MajorOptProblem_Constr2}) and the objective function (\ref{MajorOptProblem_Objective}), which are required to solve subproblem (\ref{OptSubproblem}). Then, we propose an iterative algorithm to solve (\ref{MajorOptProblem}).
	
	Applying the chain rule to (\ref{MutInf_Eqn}), the elements of the gradient vector of constraint (\ref{MajorOptProblem_Constr1}) are obtained as follows	
	\setlength{\arraycolsep}{0.0em}
	\begin{align}
	\frac{\partial I(\boldsymbol{\theta})}{\partial \boldsymbol{\theta}_i} = - \Bigg( \log_2 e &+ \int_{y} p_n(y-h_I x_i) \nonumber\\
	&\log_2 \Big( {\sum_{j} \boldsymbol{\theta}_j \, p_n(y-h_I x_j)} \Big) dy   \Bigg). 
	\label{MutInfGrad_Eqn}
	\end{align}
	\setlength{\arraycolsep}{5pt}	
	
	Furthermore, the elements of the gradient vector of the AP constraint (\ref{MajorOptProblem_Constr2}) can be calculated as follows
	
	\begin{equation}
	\frac{\partial }{\partial \boldsymbol{\theta}_i} \mathbb{E}\{x^2\} = \frac{\partial}{\partial \boldsymbol{\theta}_i} \sum_{j=0}^{S-1} \boldsymbol{\theta}_j x_j^2 = x_i^2.
	\label{APGrad_Eqn}
	\end{equation}
		
	In the following, we derive the gradient of the objective function (\ref{MajorOptProblem_Objective}) adopting the framework from \cite{Marbach2001} for maximization of a Markov chain reward with respect to an underlying parameter.
	
	It is difficult to calculate the gradient of (\ref{MajorOptProblem_Objective}) in closed-form since the limiting distribution $\pi(\xi; \boldsymbol{\theta})$ in (\ref{ExpandedAverageRewardEqn}) is not analytically tractable. 
	To deal with this problem, an algorithm was proposed in \cite{Marbach2001} to estimate the direction $\boldsymbol{f}(\boldsymbol{\theta})$ of the gradient of an average Markov chain reward by emulating the evolution of the Markov chain. 	
	The proposed algorithm may be used to update $\boldsymbol{\theta}$ in every time step. Note that, unlike in \cite{Marbach2001}, the Markov chain in Section \ref{MarkovChainSection}, has a continuous state space. 	
	
	Similar to \cite{Marbach2001}, let us introduce a vector $\boldsymbol{r}_{\xi_\mu, \xi_\nu}(\boldsymbol{\theta}) \in \mathbb{R}^S$, such that the gradient of the transition pdf $\rho(\xi_\mu, \xi_\nu;\boldsymbol{\theta})$ with respect to $\boldsymbol{\theta}$ can be written as $\nabla_{\boldsymbol{\theta}} \rho(\xi_\mu, \xi_\nu;\boldsymbol{\theta}) = \rho(\xi_\mu, \xi_\nu;\boldsymbol{\theta}) \boldsymbol{r}_{\xi_\mu, \xi_\nu}(\boldsymbol{\theta})$. 
	If a transition between $\xi_\mu$ and $\xi_\nu$ occurs with probability $\boldsymbol{\theta}_i$, then the corresponding element of vector $\boldsymbol{r}_{\xi_\mu, \xi_\nu}(\boldsymbol{\theta})$ has a non-zero value, i.e.,
	\begin{equation}		
	\label{VectorR_eqn}
	\boldsymbol{r}_{\xi_\mu, \xi_\nu}(\boldsymbol{\theta})_i = 
	\begin{cases}
	\frac{1}{\boldsymbol{\theta}_i}, &\text{if } \rho(\xi_\mu,\xi_\nu;\boldsymbol{\theta}) = \boldsymbol{\theta}_i \\
	0, &\text{otherwise}.
	\end{cases}
	\end{equation}   
		
	As in \cite{Marbach2001}, in every iteration $k$, based on the current Markov chain state $\xi^{k}$, we estimate the new direction $\boldsymbol{f}^{k+1}$ of $\nabla_{\boldsymbol{\theta}} \overline{P}(\boldsymbol{\theta}^{k+1})$ as $\boldsymbol{f}^{k+1} = \boldsymbol{f}^{k} + \boldsymbol{p}(\xi^{k}) + \big( \boldsymbol{\theta}^{k \top} \boldsymbol{p}(\xi^{k}) - \tilde{P}^{k} \big)\boldsymbol{z}^{k}$. Here, $\tilde{P}^{k}$ is the current estimate of $\overline{P}(\boldsymbol{\theta}^k)$, calculated as $\tilde{P}^{k} = \tilde{P}^{k-1} + \gamma \big( \boldsymbol{\theta}^{k-1 \top} \boldsymbol{p}(\xi^{k-1}) - \tilde{P}^{k-1} \big)$ and $\gamma$ is a positive step size. As in \cite{Marbach2001}, $\boldsymbol{z}^{k} \in \mathbb{R}^S$ is the likelihood ratio derivative vector computed as $\boldsymbol{z}^{k} = \alpha \boldsymbol{z}^{k-1} + \boldsymbol{r}_{\xi^{k-1}, \xi^{k}}(\boldsymbol{\theta}^{k-1})$, where $\boldsymbol{z}^{k} \in \mathbb{R}^S$ and $\alpha$ is a forgetting factor.

	\begin{algorithm}[h]		
		
		\SetAlgoNoLine%
		\SetKwFor{Foreach}{for each}{do}{end}
		
		Initialize: Maximum number of iterations $N_{\text{max}}$, iteration index $k=1$, and initial values $\boldsymbol{\theta}^1$, $\hat{\xi}^{1}$, $\tilde{P}^{1}$, $\boldsymbol{z}^{1}$, $\boldsymbol{f}^{1}$, $\mathcal{H}^1$, $\boldsymbol{\lambda}^1$. Matrices $\boldsymbol{\Omega}_1^*$ and $\boldsymbol{\Omega}_2^*$ are obtained in Section \ref{NN_Section}.
		
		\Repeat{\textbf{\upshape convergence or} $k=N_{\text{\upshape max}}$ }{

			1. Calculate elements of the vector $\hat{\boldsymbol{p}}(\hat{\xi}^{k})$\\
			{$\hat{\boldsymbol{p}}_n(\hat{\xi}^{k}) = \mathcal{N}_2(\hat{\xi}^{k}, h_Ex_n, \boldsymbol{\Omega}_2^*)$}\\
			2. Update direction $\boldsymbol{f}$\\
			$\boldsymbol{f}^{k+1} = \boldsymbol{f}^{k} + \hat{\boldsymbol{p}}(\hat{\xi}^{k}) + \big( \boldsymbol{\theta}^{k \top} \hat{\boldsymbol{p}}(\hat{\xi}^{k}) - \tilde{P}^{k} \big)\boldsymbol{z}^{k}$\\			
			3. Update estimate of the average reward\\
			$\tilde{P}^{k+1} = \tilde{P}^{k} + \gamma \big( \boldsymbol{\theta}^{k \top} \hat{\boldsymbol{p}}(\hat{\xi}^{k}) - \tilde{P}^{k} \big)$\\
			4. Choose a transmitted symbol $x_n$ from $\mathcal{X}$ according to the distribution $\boldsymbol{\theta}^k$\\ 
			5. Update the current state $\hat{\xi}^{k+1} = \mathcal{N}_1(\hat{\xi}^{k}, h_Ex_n, \boldsymbol{\Omega}_1^*)$\\
			6. Calculate vector $\boldsymbol{r}_{\hat{\xi}^{k}, \hat{\xi}^{k+1}}(\boldsymbol{\theta}^k)$ from (\ref{VectorR_eqn}), i.e., set all of its elements to 0, except for $\boldsymbol{r}_{\hat{\xi}^{k}, \hat{\xi}^{k+1}}(\boldsymbol{\theta}^{k})_n = \frac{1}{\boldsymbol{\theta}^k_n}$\\
			7. Update the likelihood ratio derivative $\boldsymbol{z}^{k+1} = \alpha \boldsymbol{z}^{k} + \boldsymbol{r}_{\hat{\xi}^{k}, \hat{\xi}^{k+1}}(\boldsymbol{\theta}^k)$\\
			8. Caclulate the gradient $\nabla_{\boldsymbol{\theta}} I(\boldsymbol{\theta}^k)$ by (\ref{MutInfGrad_Eqn})\\
			9. Solve (\ref{OptSubproblem}) for a given $\boldsymbol{\theta}^k$ and store the solution and corresponding Lagrangian multipliers $\{\boldsymbol{\delta}^*$, $\boldsymbol{\zeta}^{*}\}$\\
			10. Update $\boldsymbol{\theta}^{k+1} = \boldsymbol{\theta}^{k} + \boldsymbol{\delta}^{*}$ and $\boldsymbol{\lambda}^{k+1} = \boldsymbol{\zeta}^{*}$\\
			11. Update the Hessian matrix $\mathcal{H}^{k+1}$ according to (\ref{BFGS})\\
			12. Set $k = k+1$\\
		}			
		
		\caption{\strut Iterative algorithm for solving optimization problem (\ref{MajorOptProblem})}
		\label{BatchSearchAlg}
	\end{algorithm}
			
	To estimate $\boldsymbol{f}^{k+1}$, we have to be able to obtain the next Markov chain state $\xi^{k+1}$ given the current state $\xi^k$, i.e., we have to perform a random walk. To this end, in the next section, we propose a learning approach based on two DNNs to emulate the Markov chain evolution. Thereby, we choose transmitted symbol $x_n$ randomly, according to the current distribution $\boldsymbol{\theta}^k$, and estimate the next state by a DNN $\hat{\xi}^{k+1} = \mathcal{N}_1(\hat{\xi}^k, h_E x_n, \boldsymbol{\Omega}_1)$, where $\boldsymbol{\Omega}_1$ is the parameter matrix defining DNN $\mathcal{N}_1$ and $\hat{\xi}^k$ is an estimate of state $\xi^k$. Similarly, to calculate the new direction $\boldsymbol{f}^{k+1}$, we estimate the reward vector $\hat{\boldsymbol{p}}(\hat{\xi}^k)$ associated with the estimated current state $\hat{\xi}^k$ by a second DNN $\mathcal{N}_2$ with parameter matrix $\boldsymbol{\Omega}_2$, i.e., $\hat{\boldsymbol{p}}_n(\hat{\xi}^k) = \mathcal{N}_2(\hat{\xi}^k, h_E x_n, \boldsymbol{\Omega}_2)$.
		
	The proposed iterative algorithm to solve optimization problem (\ref{MajorOptProblem}) is summarized in \textbf{Algorithm \ref{BatchSearchAlg}}.
			
	\subsection{Neural Network Model for the EH Circuit}
	\label{NN_Section}
	
	In the following, we discuss a learning approach to perform the Markov chain simulation by DNNs $\hat{\xi}^{k+1} = \mathcal{N}_1(\hat{\xi}^k, h_E x_n, \boldsymbol{\Omega}_1)$ and $\hat{\boldsymbol{p}}_n(\hat{\xi}^k) = \mathcal{N}_2(\hat{\xi}^k, h_E x_n, \boldsymbol{\Omega}_2)$.
	
	As discussed in Section \ref{MarkovChainSection}, in practice, it is not possible to calculate $\xi^{k+1}$ and $\boldsymbol{p}({\xi}^k)$ analytically because of the imperfections of the EH circuit. However, due to the universal approximation theorem for DNNs \cite{Hanin2017}, estimating these values by DNNs with rectified linear unit (ReLU) activation functions in each layer is promising. Note that the estimation error depends on the network size \cite{Hanin2017}. We can achieve a high estimation precision by properly choosing the number of nodes in DNNs $\mathcal{N}_1$ and $\mathcal{N}_2$. 
	To this end, we have to train DNNs $\hat{\xi}_\nu = \mathcal{N}_1({\xi}_\mu, x_{\text{EH}}, \boldsymbol{\Omega}_1)$ and $\hat{P}'({\xi}_\mu, x_{\text{EH}}) = \mathcal{N}_2({\xi}_\mu, x_{\text{EH}}, \boldsymbol{\Omega}_2)$, where $\hat{\xi}_\nu$ is the estimate of state ${\xi}_\nu$ that follows state ${\xi}_\mu$ if $x_{\text{EH}}$ is the symbol received by the EH, and $\hat{P}'({\xi}_\mu, x_{\text{EH}})$ is the estimate of function ${P}'({\xi}_\mu, x_{\text{EH}})$. 
	In particular, if the received symbol is $x_{\text{EH}} = h_E x_n$, then $\hat{\boldsymbol{p}}_n({\xi}_\mu) = \hat{P}'({\xi}_\mu, h_E x_n) =  \mathcal{N}_2({\xi}_\mu, h_E x_n, \boldsymbol{\Omega}_2)$. Note that the training complexity and the approximation error do not depend on size of constellation set $\mathcal{X}$. 
	
	The training data for the DNNs can be obtained from a circuit simulator, such as ADS \cite{ADS2017}. For the rectenna circuit model, specified in Section \ref{EHCircuit_Section} and shown in Fig. \ref{EH_Fig}, we adopt circuit parameters similar to \cite{Morsi2019}, namely an antenna impedance $R_s = \SI{50}{\ohm}$, an SMS7630 Schottky diode, an LC matching network, fine-tuned for input signal frequency $2.45 \, \text{GHz}$ and input power value $\SI{-16}{\deci\belmilliwatt}$, a capacitor $C_L = \SI{1}{\nano\farad}$, and a load resistor $R_L = \SI{10}{\kilo\ohm}$.
	
	To train the DNNs, we randomly generate input symbols $x_{\text{EH}}$ that are independent, identically, and uniformly distributed over a space of symbols that can be feasibly received by the EH and obtain corresponding 4-tuples $\big\{ {P}'(v_L(kT), x_{\text{EH}}), v_L\big((k~+~1)T\big), v_L(kT), x_{\text{EH}}  \big\}$ using the circuit simulator. Specifically, we used 11000, 3000, and 750 4-tuples for training, validation, and testing, respectively. The training process used the Adam optimization algorithm \cite{Kingma2014} and the mean absolute percentage loss function, e.g., \cite{Myttenaere2016}. 
	
	Since the size of the DNN depends on the desired estimation error measured on the test set, we trained several networks with different numbers of layers to find the best setting. We found that the values of the mean absolute percentage error measured for the test sets for DNNs $\mathcal{N}_1({\xi}^k, x_{\text{EH}}, \boldsymbol{\Omega}_1)$ and $\mathcal{N}_2({\xi}^k, x_{\text{EH}}, \boldsymbol{\Omega}_2)$ do not decrease substantially if the size of the DNNs is increasing beyond 5 layers and 7 units per hidden layer. The network parameters $\boldsymbol{\Omega}_1^*$ and $\boldsymbol{\Omega}_2^*$ obtained after training are saved to be used for simulation of the Markov reward chain evolution in Algorithm \ref{BatchSearchAlg}.
	
	\section{Simulation Results}
	\label{SimulationResults_Section}
	
	In this section, we investigate the rate-power region of the considered SWIPT system by solving (\ref{MajorOptProblem}) with {Algorithm~\ref{BatchSearchAlg}}. 
	
	For the IR channel, we assume Rayleigh fading and a pathloss exponent of $3$. To harvest meaningful amounts of power, the TX is generally located closer to the EH than to IR. Hence, for the EH channel, we assume a line of sight and Rician fading with a Rician factor of $1$ and a pathloss exponent of $2$. 
	The distance between TX and IR is $d_{\text{IR}} = \SI{30}{\meter}$. 
	For the EH, we consider a small input power (SP) regime by setting the distance for the corresponding channel to $d_{\text{EH}} = \SI{20}{\meter}$ and a large input power (LP) regime with $d_{\text{EH}} = \SI{10}{\meter}$. 
	The distance for the LP regime was chosen such that the EH circuit may go into saturation due to the diode breakdown effect. 
	The AWGN variance at the IR is $\sigma_n^2 = \SI{-80}{\deci\belmilliwatt}$. 
	We limit the average transmitted power to $\sigma_x^2 = \SI{10}{\deci\belmilliwatt}$, and the TX peak power to $P^{\text{TX}}_{\text{max}} = \SI{52}{\dBm}$. 
	Furthermore, we adopt rectangular pulse shapes $\psi(t)$ and uniformly spaced symbols $x$, i.e., $x_k = \frac{2Ak}{S-1} - A$, where $k = {0,1,...,S-1}$ and $A$ is the TX peak signal amplitude given by $A = 10^{\frac{P^{\text{TX}}_{\text{max}}}{20}}$. For our simulations, we adopted $S = 64$. 
	In {Algorithm~\ref{BatchSearchAlg}}, we set the maximum number of iterations to $N_{\text{max}} = 4000$, the step size to $\gamma = 0.1$, and the relaxation coefficient to $\alpha = 0.1$.
	The parameters of the EH were chosen as specified in Section \ref{NN_Section}.

	\begin{figure}[!t]	
		\begin{tikzpicture}[scale=0.95]
		\begin{axis}[
		legend cell align={left},
		width=9cm,
		height=7cm,
		xlabel={$x$},		
		ylabel={Optimal input distribution, $\boldsymbol{\theta}^*$},
		xmin=-1.1, xmax=1.1,
		ymin=0, ymax=0.3,
		xtick={-1,0,1},
		xticklabels={$-A$, 0, $A$},
		ytick={0,0.1,0.2,0.3},
		legend pos=north east,
		ymajorgrids=true,
		xmajorgrids=true,
		grid style=dashed,
		legend style ={ at={(1,1)}, 
			anchor=north east, draw=black, 
			fill=white, align=left},
		samples at={0,...,20},
		ycomb,
		yticklabel style={
			/pgf/number format/fixed,
			/pgf/number format/fixed zerofill,
			/pgf/number format/precision=2
		}
		]	
				
		\addplot[color=green!50!red,mark=x] coordinates { 
			(-1, 0)
			(-1+2/63*1, 0)
			(-1+2/63*2, 0)
			(-1+2/63*3, 0)
			(-1+2/63*4, 0)
			(-1+2/63*5, 0)
			(-1+2/63*6, 0)
			(-1+2/63*7, 0)
			(-1+2/63*8, 0)
			(-1+2/63*9, 0)
			(-1+2/63*10, 0)
			(-1+2/63*11, 0)
			(-1+2/63*12, 0)
			(-1+2/63*13, 0)
			(-1+2/63*14, 0)
			(-1+2/63*15, 0.0027)
			(-1+2/63*16, 0.030)
			(-1+2/63*17, 0.02508)
			(-1+2/63*18, 0.01584)
			(-1+2/63*19, 0.01519)
			(-1+2/63*20, 0.001564)
			(-1+2/63*21, 0.0004649)
			(-1+2/63*22, 0.0)
			(-1+2/63*23, 0.0)
			(-1+2/63*24, 0)
			(-1+2/63*25, 0)
			(-1+2/63*26, 0)
			(-1+2/63*27, 0.0009386)
			(-1+2/63*28, 0.003624)
			(-1+2/63*29, 0.03102)
			(-1+2/63*30, 0.116)
			(-1+2/63*31, 0.2028)
			(-1+2/63*32, 0.2242)
			(-1+2/63*33, 0.111)
			(-1+2/63*34, 0.03279)
			(-1+2/63*35, 0.003997)
			(-1+2/63*36, 0.002033)
			(-1+2/63*37, 0)
			(-1+2/63*38, 0)
			(-1+2/63*39, 0)
			(-1+2/63*40, 0.0)
			(-1+2/63*41, 0.0)
			(-1+2/63*42, 0.0)
			(-1+2/63*43, 0.0)
			(-1+2/63*44, 0.003015)
			(-1+2/63*45, 0.01298)
			(-1+2/63*46, 0.02726)
			(-1+2/63*47, 0.02877)
			(-1+2/63*48, 0.006285)
			(-1+2/63*49, 0.001001)
			(-1+2/63*50, 0)
			(-1+2/63*51, 0)
			(-1+2/63*52, 0)
			(-1+2/63*53, 0)
			(-1+2/63*54, 0)
			(-1+2/63*55, 0)
			(-1+2/63*56, 0)
			(-1+2/63*57, 0)
			(-1+2/63*58, 0)
			(-1+2/63*59, 0)
			(-1+2/63*60, 0)
			(-1+2/63*61, 0)
			(-1+2/63*62, 0)
			(-1+2/63*63, 0)
			 };
		 
		 \addplot[color=blue,mark=square] coordinates { 
		 	(-1, 0)
		 	(-1+2/63*1, 0)
		 	(-1+2/63*2, 0)
		 	(-1+2/63*3, 0)
		 	(-1+2/63*4, 0.0026)
		 	(-1+2/63*5, 0.001963)
		 	(-1+2/63*6, 0.001984)
		 	(-1+2/63*7, 0.001301)
		 	(-1+2/63*8, 0.0019)
		 	(-1+2/63*9, 0.003153)
		 	(-1+2/63*10, 0.001932)
		 	(-1+2/63*11, 0.002508)
		 	(-1+2/63*12, 0.001364)
		 	(-1+2/63*13, 0.001511)
		 	(-1+2/63*14, 0.002161)
		 	(-1+2/63*15, 0.00177)
		 	(-1+2/63*16, 0.006345)
		 	(-1+2/63*17, 0.002931)
		 	(-1+2/63*18, 0.002839)
		 	(-1+2/63*19, 0.006576)
		 	(-1+2/63*20, 0.003336)
		 	(-1+2/63*21, 0)
		 	(-1+2/63*22, 0)
		 	(-1+2/63*23, 0)
		 	(-1+2/63*24, 0)
		 	(-1+2/63*25, 0)
		 	(-1+2/63*26, 0)
		 	(-1+2/63*27, 0.01199)
		 	(-1+2/63*28, 0.03309)
		 	(-1+2/63*29, 0.08302)
		 	(-1+2/63*30, 0.1338)
		 	(-1+2/63*31, 0.18)
		 	(-1+2/63*32, 0.1929)
		 	(-1+2/63*33, 0.135)
		 	(-1+2/63*34, 0.07748)
		 	(-1+2/63*35, 0.03372)
		 	(-1+2/63*36, 0.007639)
		 	(-1+2/63*37, 0)
		 	(-1+2/63*38, 0)
		 	(-1+2/63*39, 0)
		 	(-1+2/63*40, 0)
		 	(-1+2/63*41, 0)
		 	(-1+2/63*42, 0.003846)
		 	(-1+2/63*43, 0.003586)
		 	(-1+2/63*44, 0.002597)
		 	(-1+2/63*45, 0.00251)
		 	(-1+2/63*46, 0.004643)
		 	(-1+2/63*47, 0.003088)
		 	(-1+2/63*48, 0.001687)
		 	(-1+2/63*49, 0.001721)
		 	(-1+2/63*50, 0.002523)
		 	(-1+2/63*51, 0.00174)
		 	(-1+2/63*52, 0.002)
		 	(-1+2/63*53, 0.003)
		 	(-1+2/63*54, 0.002)
		 	(-1+2/63*55, 0.0012)
		 	(-1+2/63*56, 0.0025)
		 	(-1+2/63*57, 0.0023)
		 	(-1+2/63*58, 0.0010)
		 	(-1+2/63*59, 0)
		 	(-1+2/63*60, 0)
		 	(-1+2/63*61, 0)
		 	(-1+2/63*62, 0)
		 	(-1+2/63*63, 0)
		 };
	 
		 \addplot[color=purple,mark=x] coordinates { 
		 	(-1, 0.01454)
		 	(-1+2/63*1, 0)
		 	(-1+2/63*2, 0)
		 	(-1+2/63*3, 0)
		 	(-1+2/63*4, 0)
		 	(-1+2/63*5, 0)
		 	(-1+2/63*6, 0)
		 	(-1+2/63*7, 0)
		 	(-1+2/63*8, 0)
		 	(-1+2/63*9, 0)
		 	(-1+2/63*10, 0)
		 	(-1+2/63*11, 0)
		 	(-1+2/63*12, 0)
		 	(-1+2/63*13, 0)
		 	(-1+2/63*14, 0)
		 	(-1+2/63*15, 0)
		 	(-1+2/63*16, 0)
		 	(-1+2/63*17, 0)
		 	(-1+2/63*18, 0)
		 	(-1+2/63*19, 0)
		 	(-1+2/63*20, 0)
		 	(-1+2/63*21, 0)
		 	(-1+2/63*22, 0)
		 	(-1+2/63*23, 0)
		 	(-1+2/63*24, 0)
		 	(-1+2/63*25, 0)
		 	(-1+2/63*26, 0.0006128)
		 	(-1+2/63*27, 0.01715)
		 	(-1+2/63*28, 0.08099)
		 	(-1+2/63*29, 0.1549)
		 	(-1+2/63*30, 0.1678)
		 	(-1+2/63*31, 0.1969)
		 	(-1+2/63*32, 0.1714)
		 	(-1+2/63*33, 0.08657)
		 	(-1+2/63*34, 0.04967)
		 	(-1+2/63*35, 0.01172)
		 	(-1+2/63*36, 0.003725)
		 	(-1+2/63*37, 0)
		 	(-1+2/63*38, 0)
		 	(-1+2/63*39, 0)
		 	(-1+2/63*40, 0)
		 	(-1+2/63*41, 0)
		 	(-1+2/63*42, 0)
		 	(-1+2/63*43, 0)
		 	(-1+2/63*44, 0)
		 	(-1+2/63*45, 0)
		 	(-1+2/63*46, 0)
		 	(-1+2/63*47, 0)
		 	(-1+2/63*48, 0)
		 	(-1+2/63*49, 0)
		 	(-1+2/63*50, 0)
		 	(-1+2/63*51, 0)
		 	(-1+2/63*52, 0)
		 	(-1+2/63*53, 0)
		 	(-1+2/63*54, 0)
		 	(-1+2/63*55, 0)
		 	(-1+2/63*56, 0)
		 	(-1+2/63*57, 0)
		 	(-1+2/63*58, 0)
		 	(-1+2/63*59, 0)
		 	(-1+2/63*60, 0)
		 	(-1+2/63*61, 0)
		 	(-1+2/63*62, 0)
		 	(-1+2/63*63, 0.01882)
		 };
	 
		 \addplot[color=orange,mark=square] coordinates { 
		 	(-1, 0.01465)
		 	(-1+2/63*1, 0)
		 	(-1+2/63*2, 0)
		 	(-1+2/63*3, 0)
		 	(-1+2/63*4, 0)
		 	(-1+2/63*5, 0)
		 	(-1+2/63*6, 0)
		 	(-1+2/63*7, 0)
		 	(-1+2/63*8, 0)
		 	(-1+2/63*9, 0)
		 	(-1+2/63*10, 0)
		 	(-1+2/63*11, 0)
		 	(-1+2/63*12, 0)
		 	(-1+2/63*13, 0)
		 	(-1+2/63*14, 0)
		 	(-1+2/63*15, 0)
		 	(-1+2/63*16, 0)
		 	(-1+2/63*17, 0)
		 	(-1+2/63*18, 0)
		 	(-1+2/63*19, 0)
		 	(-1+2/63*20, 0)
		 	(-1+2/63*21, 0)
		 	(-1+2/63*22, 0)
		 	(-1+2/63*23, 0)
		 	(-1+2/63*24, 0.001704)
		 	(-1+2/63*25, 0.0048)
		 	(-1+2/63*26, 0.0126)
		 	(-1+2/63*27, 0.02)
		 	(-1+2/63*28, 0.02395)
		 	(-1+2/63*29, 0.07235)
		 	(-1+2/63*30, 0.1263)
		 	(-1+2/63*31, 0.1676)
		 	(-1+2/63*32, 0.1727)
		 	(-1+2/63*33, 0.1355)
		 	(-1+2/63*34, 0.1095)
		 	(-1+2/63*35, 0.06966)
		 	(-1+2/63*36, 0.03251)
		 	(-1+2/63*37, 0.01173)
		 	(-1+2/63*38, 0.002494)
		 	(-1+2/63*39, 0)
		 	(-1+2/63*40, 0)
		 	(-1+2/63*41, 0)
		 	(-1+2/63*42, 0)
		 	(-1+2/63*43, 0)
		 	(-1+2/63*44, 0)
		 	(-1+2/63*45, 0)
		 	(-1+2/63*46, 0)
		 	(-1+2/63*47, 0)
		 	(-1+2/63*48, 0)
		 	(-1+2/63*49, 0)
		 	(-1+2/63*50, 0)
		 	(-1+2/63*51, 0)
		 	(-1+2/63*52, 0)
		 	(-1+2/63*53, 0)
		 	(-1+2/63*54, 0)
		 	(-1+2/63*55, 0)
		 	(-1+2/63*56, 0)
		 	(-1+2/63*57, 0)
		 	(-1+2/63*58, 0)
		 	(-1+2/63*59, 0)
		 	(-1+2/63*60, 0)
		 	(-1+2/63*61, 0)
		 	(-1+2/63*62, 0)
		 	(-1+2/63*63, 0.01882)
		 };

		\legend{$T=\SI{100}{\micro\second}$ (LP), $T=\SI{1}{\micro\second}$ (LP), $T=\SI{100}{\micro\second}$ (SP),  $T=\SI{1}{\micro\second}$ (SP),}	
		\end{axis}

		\end{tikzpicture}		
		\caption{Optimal input distribution as solution of (\ref{MajorOptProblem}) with $I_{\text{req}} = \SI{3}{\frac{\bit}{symbol}}$ for $T= \SI{1}{\micro\second}$ and $T= \SI{100}{\micro\second}$ in SP and LP regimes.}
		\label{Distributions_Fig}
		\centering
	\end{figure}
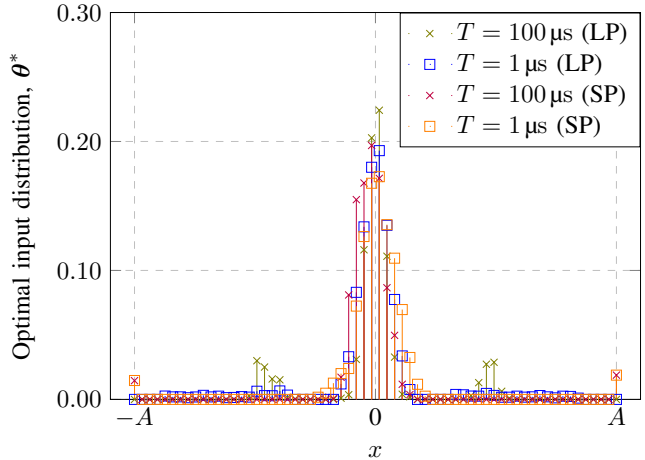

	In Fig.~\ref{Distributions_Fig}, we show the optimal input distribution $\boldsymbol{\theta}^*$ obtained by solving optimization problem (\ref{MajorOptProblem}) with {Algorithm \ref{BatchSearchAlg}} for a required mutual information of $I_{\text{req}} = \SI{3}{\frac{\bit}{symbol}}$ and a given realization of the Rayleigh and Rician fading. 
	In particular, we show the optimal input distributions for small ($T = \SI{1}{\micro\second}$) and large ($T = \SI{100}{\micro\second}$) symbol durations for the SP and LP regimes.
	We observe that the optimal input distribution does not depend much on the symbol duration $T$ in the SP regime, where it is optimal to allocate a small probability to symbols having the maximum amplitude $A$, as even for this large amplitude, saturation is not reached. 
	In the LP regime, it is optimal to limit the symbol amplitude to a smaller value for $T = \SI{100}{\micro\second}$ to avoid driving the EH circuit into saturation. 
	In contrast, if the symbol duration is small, small but non-zero probabilities are allocated to symbols with high amplitudes even in the LP regime. 
	In fact, since the capacitor $C_L$ in the EH circuit cannot be fully charged within one symbol interval if the symbol duration is small, the saturation behavior of the EH depends on $T$ as well. 
	Hence, in the LP regime, the optimal input distribution depends on the symbol duration.

	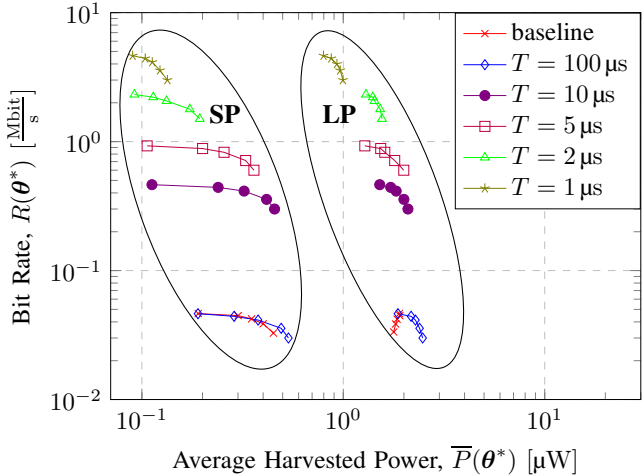
\begin{figure}[!t]	
		\begin{tikzpicture}[scale=0.95]
		\begin{axis}[
		legend cell align={left},
		ymode=log,
		xmode=log,
		width=9cm,
		height=7cm,
		xlabel={Average Harvested Power, $\overline{P}(\boldsymbol{\theta}^*)$ $[\SI{}{\micro\watt}]$},		
		ylabel={Bit Rate, $R(\boldsymbol{\theta}^*)$ $[ \frac{ \SI{}{\mega\bit}}{ \SI{}{\second} } ] $},
		xmin=0.07, xmax=30,
		ymin=0.01, ymax=10,
		xtick={0.1,1,10},
		ytick={0.01,0.1,1,10},
		legend pos=north east,
		ymajorgrids=true,
		xmajorgrids=true,
		grid style=dashed,
		legend style ={ at={(1,1)}, 
			anchor=north east, draw=black, 
			fill=white, align=left},
		]	
		\addplot[color=red,mark=x] coordinates { 
			(0.1905, 4.64 /100) 
			(0.3006, 4.472 /100) 
			(0.3504, 4.22 /100) 
			(0.4002, 3.865 /100) 
			(0.45, 3.274 /100) };		
		\addplot[color=blue,mark=diamond] coordinates { 
			(0.19, 4.63 /100) 
			(0.2871, 4.416 /100) 
			(0.3777, 4.132 /100) 
			(0.4917, 3.566 /100) 
			(0.5342, 3 /100) };
		\addplot[color=violet,mark=*] coordinates { 
			(0.4558, 3 /10) 
			(0.416, 3.565 /10) 
			(0.3218, 4.132 /10) 
			(0.2391, 4.416 /10) 
			(0.1122, 4.634 /10)  };
		\addplot[color=purple,mark=square] coordinates {
			(0.3601, 3 /5) 
			(0.3276, 3.565 /5) 
			(0.2558, 4.132 /5) 
			(0.2002, 4.416 /5) 
			(0.1062, 4.634 /5)	};	
		\addplot[color=green,mark=triangle] coordinates { 
			(0.1939, 3 /2) 
			(0.1730, 3.565 /2) 
			(0.1326, 4.132 /2) 
			(0.1139, 4.417 /2) 
			(0.0918, 4.634 /2)  };		
		\addplot[color=green!50!red,mark=star] coordinates { 
			(0.1339, 3 /1) 
			(0.1230, 3.565 /1) 
			(0.1126, 4.132 /1) 
			(0.1039, 4.417 /1) 
			(0.0900, 4.634 /1)  };

		\addplot[color=red,mark=x] coordinates { 
			(1.9, 4.64 /100) 
			(1.91, 4.468 /100) 
			(1.85, 4.197 /100) 
			(1.82, 3.854 /100) 
			(1.78, 3.341 /100) };		
		\addplot[color=blue,mark=diamond] coordinates { 
			(1.868, 4.63 /100) 
			(2.171, 4.416 /100) 
			(2.282, 4.132 /100) 
			(2.393, 3.566 /100) 
			(2.478, 3 /100) };
		\addplot[color=violet,mark=*] coordinates { 
			(2.094, 3 /10) 
			(2, 3.565 /10) 
			(1.831, 4.132 /10) 
			(1.72, 4.416 /10) 
			(1.517, 4.634 /10)  };
		\addplot[color=purple,mark=square] coordinates {
			(1.997, 3 /5) 
			(1.8, 3.565 /5) 
			(1.607, 4.132 /5) 
			(1.534, 4.416 /5) 
			(1.279, 4.634 /5)	};	
		\addplot[color=green,mark=triangle] coordinates { 
			(1.557, 3 /2) 
			(1.525, 3.565 /2) 
			(1.430, 4.132 /2) 
			(1.402, 4.417 /2) 
			(1.296, 4.634 /2)  };		
		\addplot[color=green!50!red,mark=star] coordinates { 
			(0.9962, 3 /1) 
			(0.9653, 3.565 /1) 
			(0.93, 3.992 /1) 
			(0.872, 4.417 /1) 
			(0.798, 4.634 /1)  };	
		
		\legend{baseline, {$T=\SI{100}{\micro\second}$}, $T=\SI{10}{\micro\second}$, $T=\SI{5}{\micro\second}$, $T=\SI{2}{\micro\second}$, $T=\SI{1}{\micro\second}$}	
		\end{axis}	
		\draw[rotate around={20:(1.4,2.8)}] (1.4,2.8) ellipse (1 and 2.5);	
		\draw (1.6,4) node {\normalsize \textbf{SP}};

		\draw[rotate around={20:(3.8,2.8)}] (3.8,2.8) ellipse (0.8 and 2.5);	
		\draw (3.2,4) node {\normalsize \textbf{LP}};
		\end{tikzpicture}		
		\caption{Rate-power region for the SP and LP regimes and different symbol durations $T$.}
		\label{MI_AHP_Fig}
		\centering
	\end{figure}

	In Fig.~\ref{MI_AHP_Fig}, we show the boundaries of the rate-power region obtained by solving optimization problem~(\ref{MajorOptProblem}) for different required mutual information values $I_{\text{req}}$ with {Algorithm~\ref{BatchSearchAlg}}.
	Results for different symbol durations and for the SP and LP regimes are depicted. 
	The simulation results were averaged over 1000 channel realizations. 
	The average harvested power $\overline{P}(\boldsymbol{\theta}^*)$ was obtained by ADS circuit simulations, whereas the bit rate was calculated as $R(\boldsymbol{\theta}^*) = \frac{I(\boldsymbol{\theta}^*)}{T}$. 
				
	As baseline scheme, we adopt the input distribution proposed in \cite{Morsi2019}. In \cite{Morsi2019}, the memory effect of the EH was neglected, i.e., an infinitely large symbol duration $T$ was assumed. For the sake of comparison, we normalize the obtained mutual information to a sufficiently large value of $T$, i.e., $T = \SI{100}{\micro\second}$.
	Moreover, in \cite{Morsi2019}, the authors assumed perfect matching between antenna and rectifier in the EH circuit for every value of the input signal power.
	We observe that the baseline scheme achieves the same performance as the proposed scheme for the case of maximum information rate, i.e., when the value $I_\text{req}$ in (\ref{MajorOptProblem_Constr1}) is large.
	However, for smaller values of $I_\text{req}$, we observe that while, in the SP regime, the baseline rate-power region is only slightly worse than the rate-power region obtained with {Algorithm \ref{BatchSearchAlg}, in the LP regime, the proposed scheme outperforms the baseline scheme significantly. This gain is due to the more accurate modeling of the EH circuit non-idealities, such as imperfect matching and diode breakdown, enabled by DNNs.
		
	For the proposed scheme, we observe that a smaller symbol duration generally leads to a higher bit rate $R$. Additionally, we observe that, for any value of symbol duration $T$, the average harvested power in the LP regime is larger than the one in the SP regime. However, in both input power regimes, decreasing the symbol duration leads to a significant reduction of the average power that can be harvested by the EH. 
	Moreover, we observe that for small symbol duration values, e.g., $T = \SI{1}{\micro\second}$, for both power regimes, the average harvested power saturates at a low value and cannot be improved much by relaxing the constraint $I_\text{req}$.
	Fig.~\ref{MI_AHP_Fig} reveals that the rate-power region of the considered SWIPT system is affected by both the symbol duration and the input power value in the EH since the rectenna memory has a significant influence on the harvested power.

	\section{Conclusion}
	\label{Conclusion_Section}
	
	In this paper, we considered SWIPT systems employing nonlinear EH circuits with memory. We modeled the memory of the EH circuit by a Markov reward chain. Additionally, we proposed a learning approach to model the imperfections of the EH circuit. We formulated and solved an optimization problem to determine the trade-off between the achievable information rate and the harvested power. Our simulation results revealed that, for high EH input power levels, the optimal distribution depends on the symbol duration. Furthermore, our results showed that while shorter symbol durations increase the bit rate, they have a negative effect on the harvested power.

	\bibliographystyle{IEEEtran}
	\bibliography{Paper}
	
\end{document}